\documentclass{PoS}

\title{
    \begin{picture}(0,0)(0,0)%
    \put(350,55){\makebox(0,0)[l]{\textnormal{\normalsize RIKEN-QHP-386}}}%
    \end{picture}
HAL QCD method and Nucleon-Omega interaction with physical quark masses}

\ShortTitle{HAL QCD method and Nucleon-Omega interaction with physical quark masses}

\author{\speaker{Takumi Iritani}\\
  Theoretical Research Division, Nishina Center, RIKEN, Wako 351-0198, Japan \\
  E-mail: \email{takumi.iritani@riken.jp}}

\author{for HAL QCD Collaboration\\
  \includegraphics[width=0.35\textwidth]{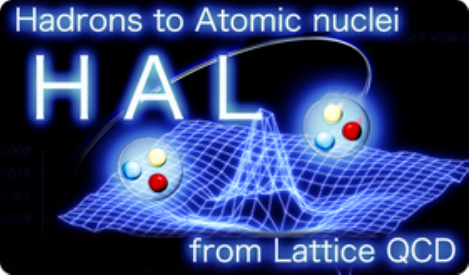}
}

\abstract{
  In lattice QCD, both direct method and HAL QCD method are used to investigate the two-baryon systems.
  We show that due to the contamination of the scattering excited states,
  it is challenging to measure the eigenenergy from the temporal correlation in the direct method,
  while the HAL QCD method can extract the information of the interaction from both scattering states and ground state 
  by using the spatial correlation.
  We examine the systematic uncertainty of the derivative expansion in the HAL QCD method, 
  which is found to be well under control at the low energies.
  By using the time-dependent HAL QCD method,
  we study the nucleon($N$)-Omega($\Omega$) system in the $^5$S$_2$ channel with almost physical 
  quark masses at $m_\pi \simeq 146$~MeV.
  We find the interaction is attractive at all distances,
  which produces a quasi-bound state with the binding energy 1.54(0.30)($^{+0.04}_{-0.10}$)~MeV.
  We also consider the extra Coulomb interaction in the $p\Omega^{-}$($^5$S$_2$) system,
  whose binding energy becomes 2.46(0.34)($^{+0.04}_{-0.01}$)~MeV.
  $N\Omega$($^5$S$_2$) dibaryon could be searched through two-particle correlations
  in the heavy ion collision experiments.
}

\FullConference{The 36th Annual International Symposium on Lattice Field Theory - LATTICE2018\\
		22-28 July, 2018\\
		Michigan State University, East Lansing, Michigan, USA.}

\begin{document}

\section{Introduction}

The search for the dibaryons is one of the long-standing problems in hadron physics.
The most famous candidate is the $H$-dibaryon($uuddss$), which is predicted by the MIT bag model in 1977~\cite{Jaffe:1976yi}.
Recently, the model-independent studies of the $H$-dibaryon 
are reported from lattice QCD calculations~\cite{Inoue:2011ai, Sasaki:2018mzh, Francis:2018qch}.
Due to the absence of the Pauli blocking effect,
another promising candidate for the dibaryon is $N\Omega$($uudsss$ or $uddsss$),
which is claimed to be bound
from the studies based on the phenomenological models~\cite{
  Goldman:1987ma, Oka:1988yq, Li:1999bc, Pang:2003ty, Zhu:2015sna, Huang:2015yza, Sekihara:2018tsb}.
This dibaryon is also reported
from the (2+1)-flavor lattice QCD study for the heavier pion mass at $m_\pi \simeq 875$~MeV~\cite{Etminan:2014tya}.

In order to clarify the existence of the dibaryon, the reliable calculation from lattice QCD is important.
However, in the previous studies for the two-baryon systems for heavier pion masses,
inconsistent conclusions are reported from two lattice QCD approaches~\cite{Davoudi:2017ddj, Ishii:2006ec, Aoki:2012tk}.
In this work, we discuss the fundamental difficulty in the two-baryon systems from the direct method
by using the temporal correlation
due to the contamination of the scattering states,
while the HAL QCD method is free from such an issue by using the spatial correlation~\cite{HALQCD:2012aa, Iritani:2016jie, Iritani:2018zbt}.
We also show the systematic uncertainty of the derivative expansion in the HAL QCD method is well under control.
By using the HAL QCD method, we discuss the $N\Omega$ system in the $^5$S$_2$ channel 
with almost physical quark masses at $m_\pi \simeq 146$~MeV~\cite{Iritani:2018sra}.

\section{Two-Baryon Systems from Lattice QCD}

\subsection{Direct method and the pseudo plateaux problem}

In the direct method, the energy eigenvalue of the two-baryon system is extracted from the temporal correlation.
In practical lattice QCD calculations,
the energy shift $\Delta E_L = E_\mathrm{BB}^{L} - 2m_B$ at a finite box with
the spatial extension $L$ is obtained by the plateau of the effective energy shift
$\Delta E_\mathrm{BB}^\mathrm{eff}(t)$, which is defined by
\begin{equation}
  \Delta E_\mathrm{BB}^\mathrm{eff}(t)
  \equiv \frac{1}{a} \log \frac{R_\mathrm{BB}(t)}{R_\mathrm{BB}(t+a)}
  \label{eq:Eeff}
\end{equation}
where $R_\mathrm{BB}(t) \equiv C_\mathrm{BB}(t)/\{C_\mathrm{B}(t)\}^2$
with the two-baryon (single baryon) correlator $C_\mathrm{BB}(t)$ ($C_\mathrm{B}(t)$)
and a lattice spacing $a$. 

One of the problems in the multi-baryon systems is the signal to noise ratio,
which becomes exponentially worse
as $S(t)/N(t) \sim \exp\left[ - A(m_B - (3/2)m_M)t\right]$,
where $A$ is the baryon number, $m_B$ and $m_M$ are the baryon and meson masses, respectively.
Furthermore, the contamination of the elastic excited states can be a severe problem.
The energy gap of the scattering states is proportional to $\mathcal{O}(1/L^2)$,
which is much smaller than the inelastic gap $\mathcal{O}(\Lambda_\mathrm{QCD})$.
It means the ground state saturation of the two-baryon system
requires a large Euclidean time than that of the single particle system.

\begin{figure}[h]
  \centering
  \includegraphics[width=0.47\textwidth,clip]{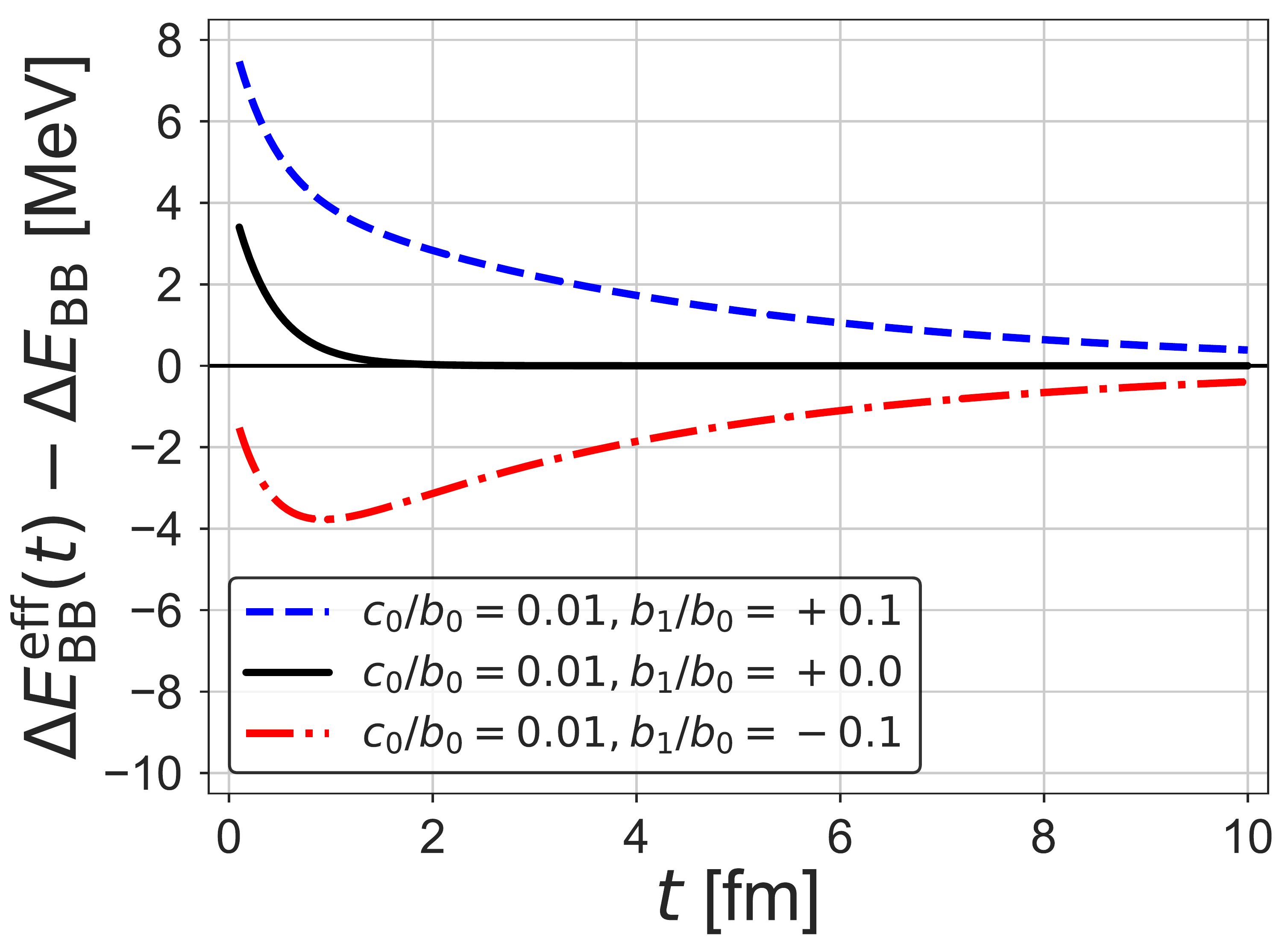}
  \includegraphics[width=0.47\textwidth,clip]{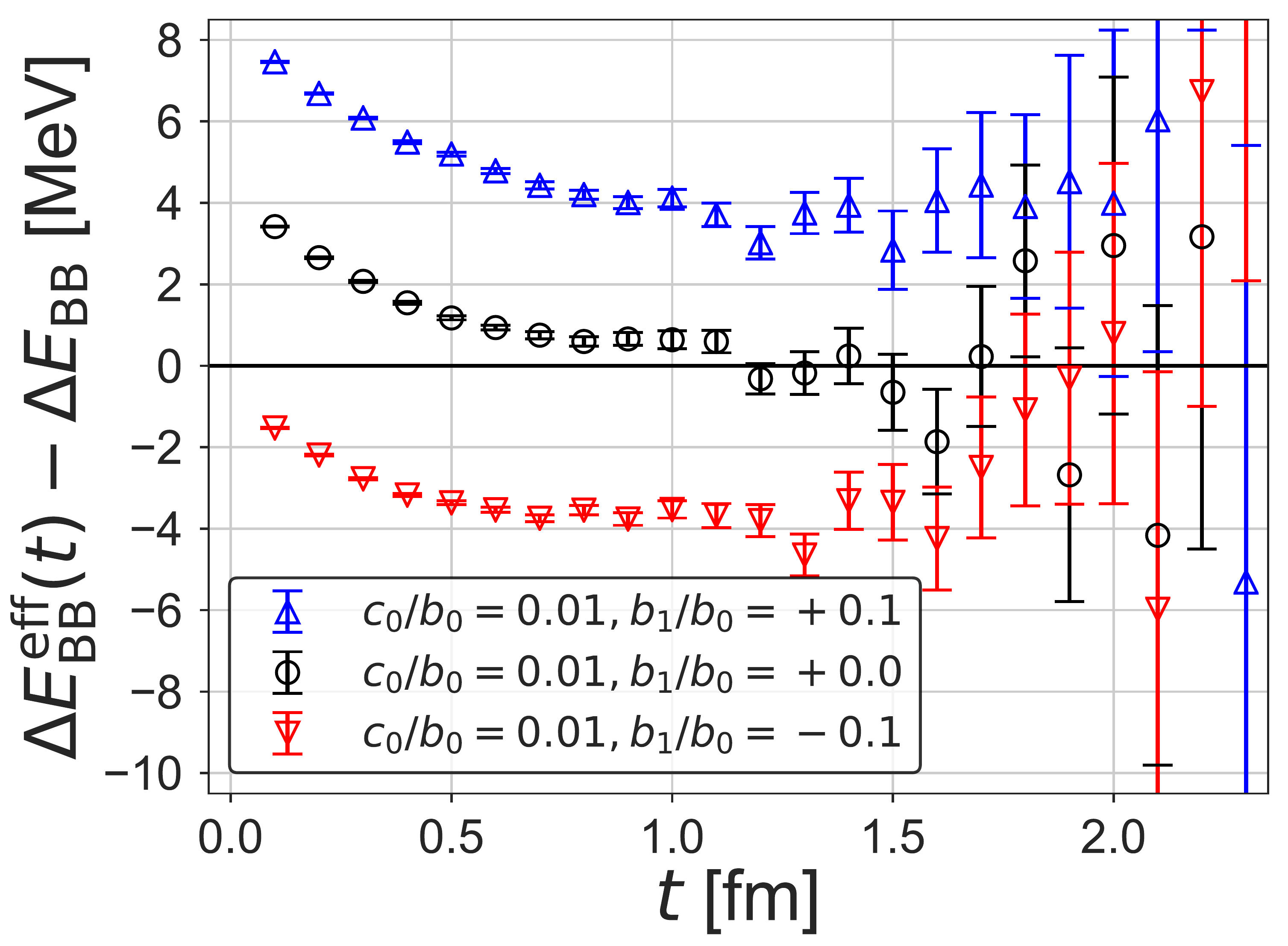}
  \caption{(Left) The effective energy shift of the mock-up data. 
    (Right) A mock-up data with fluctuations.
  \label{fig:mock}}
\end{figure}

To demonstrate this point~\cite{Iritani:2016jie},
we consider a mock-up correlator as
\begin{equation}
  R(t) = b_0 e^{-\Delta E_\mathrm{BB}t} + b_1 e^{-(\delta E_\mathrm{el}+\Delta E_\mathrm{BB})t}
  +c_0 e^{-(\delta E_\mathrm{inel}+\Delta E_\mathrm{BB})t},
  \label{eq:mock_up}
\end{equation}
with the energy shift $\Delta E_\mathrm{BB} = E_\mathrm{BB} - 2m_\mathrm{B}$,
and the energy gap of the elastic(inelastic) excited state $\delta E_\mathrm{el}$($\delta E_\mathrm{inel}$).
Here, we adopt $\delta E_\mathrm{el} = 50$~MeV and $\delta E_\mathrm{inel} = 500$~MeV,
which are typical scales of the current lattice QCD simulations.
Fig.~\ref{fig:mock}~(Left)
shows $\Delta E_\mathrm{BB}^\mathrm{eff}(t) - \Delta E_\mathrm{BB}$
for $c_0/b_0 = 0.01$ and $b_1/b_0 = 0.1$, $0$, $-0.1$.
The inelastic state becomes negligible around 1~fm,
while $\mathcal{O}(10)$~fm of the Euclidean time is required for 
the ground state saturation with the elastic excited state. 

The effective energy shifts with fluctuations are shown in Fig.~\ref{fig:mock}~(Right).
There are plateau-like structures around $t \sim 1$~fm.
However, these are incorrect signals except black circles ($b_1/b_0 = 0$).
It shows the plateau-like behavior
cannot guarantee the ground state saturation at all.

These pseudo plateaux cast a question on the previous studies 
of two-nucleon systems by the direct method~\cite{Davoudi:2017ddj},
which depend on the naive plateaux fitting without employing a variational method.
In Ref.~\cite{Iritani:2017rlk,Aoki:2017byw}, we discuss the normality of these results
based on the L\"uscher's finite volume formula~\cite{Luscher:1991},
and clarify all of them show anomalous behaviors, for example,
the parameters of the effective range expansion are found to be singular,
the S-matrix pole has the unphysical residue.
These symptoms imply the misidentification of the eigenenergies.

\subsection{HAL QCD method and the convergence of the derivative expansion}

The time-dependent HAL QCD method
extracts the information of the interaction 
by using all scattering states below the inelastic threshold~\cite{HALQCD:2012aa}.
In this method, the energy-independent and non-local potential $U(\vec{r}, \vec{r'})$ is given by
\begin{equation}
  \left[ 
  - H_0
  - \frac{\partial}{\partial t} 
  + \frac{1}{4m_B}\frac{\partial^2}{\partial t^2}
\right]R(\vec{r},t)
= \int d\vec{r^\prime} U(\vec{r},\vec{r^\prime})R(\vec{r^\prime},t).
\end{equation}
Here the $R$-correlator is defined as
\begin{equation}
  R(\vec{r},t) \equiv 
  \left\langle 0 | T\{B(\vec{x}+\vec{r},t)B(\vec{x},t)\overline{\mathcal{J}}(0) | 0 \right\rangle
    /\{C_{B}(t)\}^2
    = \sum_n A_n \psi^{W_n}(\vec{r}) e^{-\Delta W_n t}
    + \mathcal{O}(e^{-\Delta W_\mathrm{th}t})
\end{equation}
with a source operator $\mathcal{J}$,
the Nambu-Bethe-Salpeter wave function $\psi^{W_n}(r)$, 
$\Delta W_n = W_n - 2m_B$ with $n$-th energy eigenvalue $W_n$,
and the inelastic threshold 
$\Delta W_\mathrm{th} = W_\mathrm{th} - 2m_B$.
%
For the spin-singlet channel, 
the potential in the leading order (LO) analysis 
of the velocity expansion 
$U(\vec{r}, \vec{r'}) = \sum_n V_n(\vec{r}) \nabla^n \delta(\vec{r}-\vec{r'})$
is given by
\begin{equation}
  V_0^\mathrm{LO}(r) = - \frac{H_0 R(\vec{r},t)}{R(\vec{r},t)}
  - \frac{(\partial/\partial t)R(\vec{r}, t)}{R(\vec{r},t)}
  + \frac{1}{4m_B} \frac{(\partial^2/\partial t^2)R(\vec{r}, t)}{R(\vec{r},t)}.
  \label{eq:HAL_LO}
\end{equation}
To confirm its convergence,
we consider the next-to-next-to-leading order (N$^2$LO) analysis as $U(\vec{r},\vec{r'}) \simeq \{ V_0^\mathrm{N^2LO}(r)
+ V_2^\mathrm{N^2LO}(r)\nabla^2 \}\delta(\vec{r}-\vec{r'})$.
The relation among these potentials is given by
\begin{equation}
  V_0^\mathrm{LO}(r) = V_0^\mathrm{N^2LO}(r) + V_2^\mathrm{N^2LO}(r)\frac{\nabla^2 R(\vec{r}, t)}{R(\vec{r}, t)},
  \label{eq:LO_N2LO}
\end{equation}
which means the N$^2$LO correction in $V_0^\mathrm{LO}$ depends on both 
$V_2^\mathrm{N^2LO}$ and the $R$-correlator.

\begin{figure}
  \centering
  \includegraphics[width=0.47\textwidth,clip]{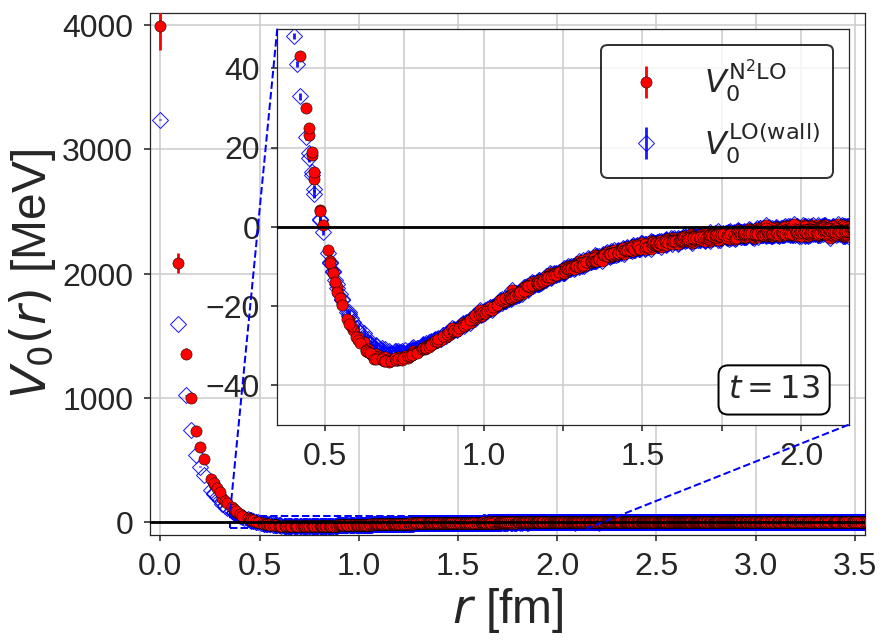}
  \includegraphics[width=0.45\textwidth,clip]{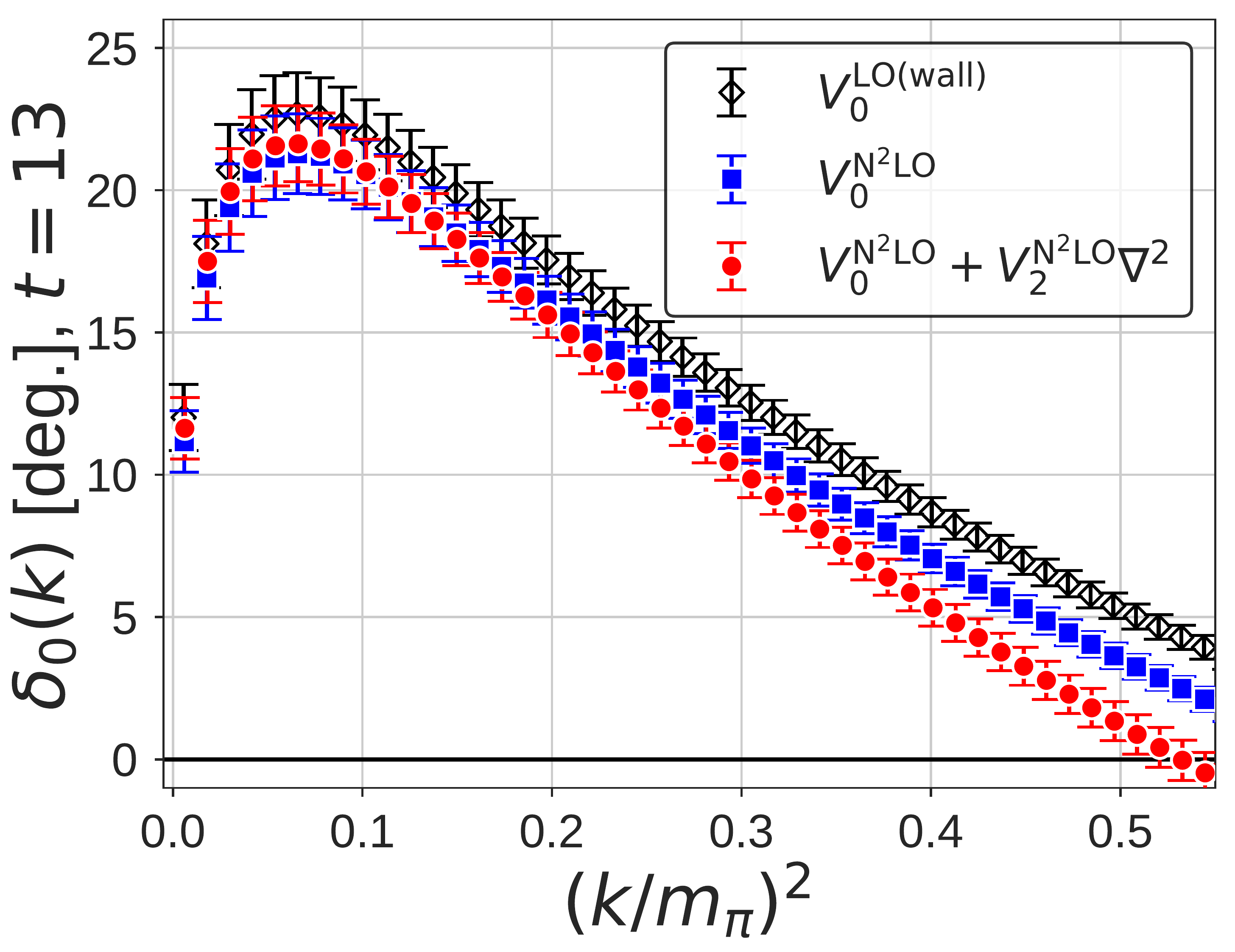}
  \caption{
    \label{fig:hal_pot}
    (Left) 
    The leading order potential of $\Xi\Xi$($^1$S$_0$) at $m_\pi = 510$~MeV.
    (Right) 
    The S-wave scattering phase shift as a function of $(k/m_\pi)^2$ from
    $V_0^\mathrm{LO(wall)}$, $V_0^\mathrm{N^2LO}$ and $V_0^\mathrm{N^2LO} + V_2^\mathrm{N^2LO}\nabla^2$.
    Both figures are taken from~\cite{Iritani:2018zbt}.
  }
\end{figure}

Fig.~\ref{fig:hal_pot}~(Left) shows $V_0$ potential
of $\Xi\Xi$($^1$S$_0$) at $m_\pi = 510$~MeV
from the LO analysis by using wall-type quark source and the N$^2$LO analysis.
The S-wave scattering phase shifts from the LO and the N$^2$LO analyses are shown in Fig.~\ref{fig:hal_pot}~(Right).
As shown in these figures,
the LO analysis from the wall source works well at the low energies,
while the N$^2$LO correction appears at high energies.
These results imply that the systematic uncertainties of the derivative expansion are well under control at the low energies~\cite{Iritani:2018zbt}.

\section{$N\Omega$ system at almost physical quark masses}

Next we apply the HAL QCD method to the $N\Omega$ system in the $^5$S$_2$ channel
at almost physical masses ($m_\pi \simeq 146$~MeV and $m_K \simeq 525$~MeV)~\cite{Iritani:2018sra}.
The lattice volume is $96^4$ with a lattice cutoff $a^{-1} = 2.333$~GeV~\cite{Ishikawa:2015rho}.
We employ the wall-type quark source with the Coulomb gauge fixing.
Total number of the measurements is 119,232,\footnote{The statistics are slightly improved
from the result at the conference.}
and the statistical errors are estimated by the jack-knife sampling.
The masses of a nucleon and $\Omega$
are 954.7(2.7)~MeV and 1711.5(1.0)~MeV, respectively, which are slightly heavier than the physical values.

In this work, we consider a single channel potential of the $N\Omega$($^5$S$_2$)
in the LO analysis of the derivative expansion.
Strictly speaking, $N\Omega$($^5$S$_2$) decays into the D-wave states of 
$\Sigma\Xi$ and $\Lambda\Xi$. 
We assume the coupling to these states is kinematically suppressed\footnote{
  The effect of these coupling channels is found to be small from the phenomenological study in Ref.~\cite{Sekihara:2018tsb},
  nevertheless, the coupled channel analysis would be required to confirm this assumption in the future~\cite{Sasaki:2018mzh}.
}.

\begin{figure}
  \centering
  \includegraphics[width=0.47\textwidth,clip]{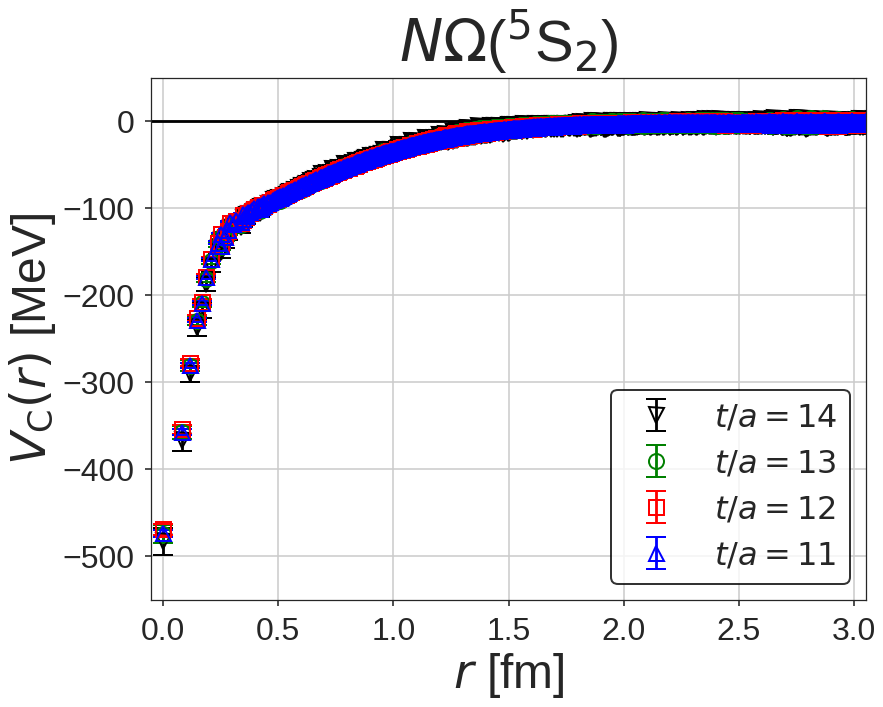}
  \includegraphics[width=0.47\textwidth,clip]{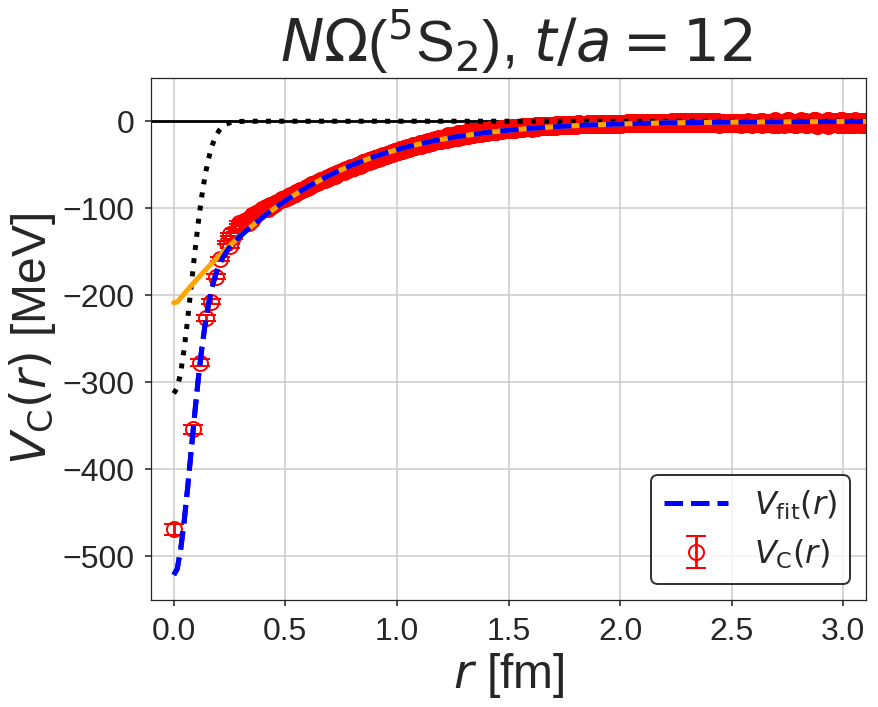}
  \caption{
    \label{fig:NOmegaPot}
    (Left) The central potential $V_c(r)$ of the $N\Omega(^5$S$_2)$ at
    $t/a = 11-14$.
    (Right) The result of the fitting of $V_\mathrm{C}(r)$ (red circles)
    by using Eq.~(\ref{eq:fit_func}) at $t/a = 12$ (blue dashed line).
    The black dotted and the orange solid lines correspond to the first and
    the second term in Eq.~(\ref{eq:fit_func}), respectively.
    Both figures are taken from~\cite{Iritani:2018sra}.
  }
\end{figure}

Fig.~\ref{fig:NOmegaPot}~(Left) shows the central potential $V_\mathrm{C}(r)$ at $t/a = 11- 14$.
These results are consistent with each other within the statistical errors.
It indicates the smallness of the coupling to the D-wave octet-octet systems.
$V_\mathrm{C}(r)$ is attractive in all distances,
which is qualitatively the same as the previous study at heavier quark masses~\cite{Etminan:2014tya}.

In order to calculate the scattering phase shifts and the binding energy, 
we fit the potential by using
Gaussian + (Yukawa)$^2$ with a form factor~\cite{Etminan:2014tya} as
\begin{equation}
  V_\mathrm{fit}(r) = c_0 e^{-c_1r^2} + c_2\left( 1 - e^{-c_3 r^2} \right)^n
  \left(\frac{e^{-m_\pi r}}{r} \right)^2.
  \label{eq:fit_func}
\end{equation}
We find that $n = 1$ with $m_\pi = 146$~MeV
works well as shown in Fig.~\ref{fig:NOmegaPot}~(Right).
The details of the analyses and parameters are summarized in Ref.~\cite{Iritani:2018sra}.

\begin{figure}
  \centering
  \includegraphics[width=0.47\textwidth,clip]{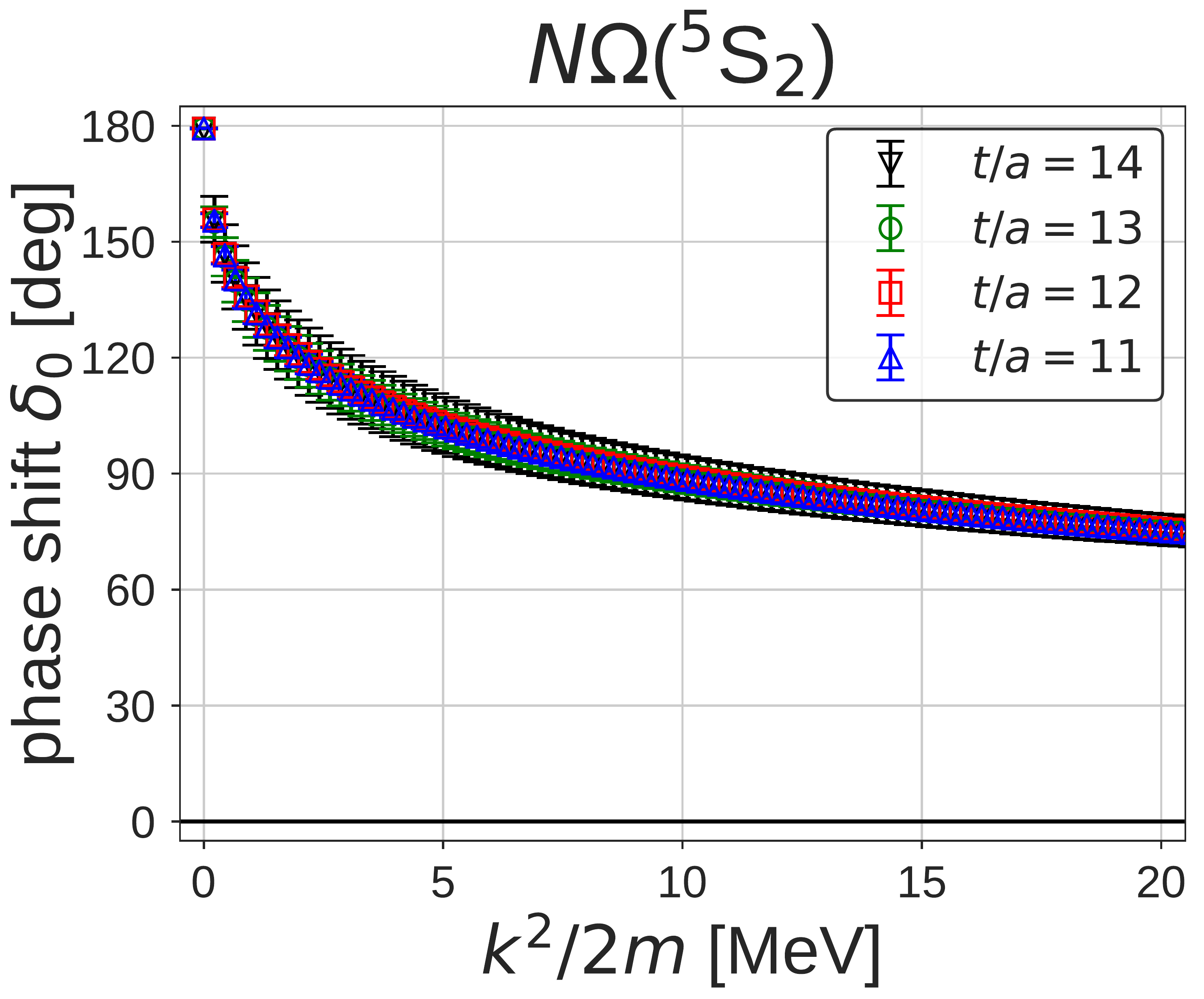}
  \includegraphics[width=0.47\textwidth,clip]{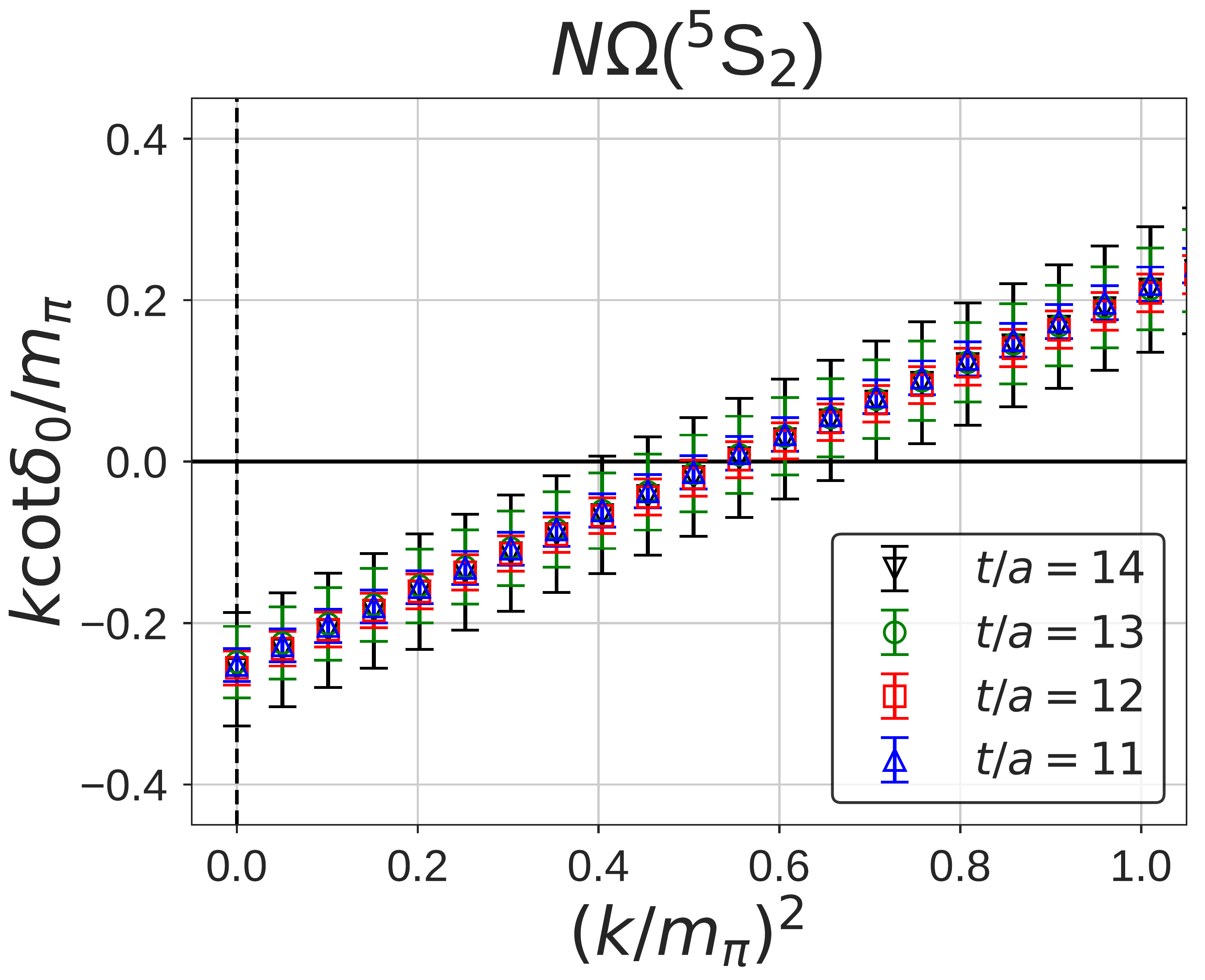}
  \caption{
    \label{fig:NOmegaPhaseShift}
    (Left) The S-wave scattering phase shifts $\delta_0$ as a function of the kinetic energy
    from $t/a = 11-14$.
    (Right) $k\cot\delta_0/m_\pi$ as a function of $(k/m_\pi)^2$ from $t/a = 11-14$.
    Both figures are taken from~\cite{Iritani:2018sra}.
  }
\end{figure}

Fig.~\ref{fig:NOmegaPhaseShift}~(Left) shows
the S-wave scattering phase shifts $\delta_0$ as a function of the kinetic energy.
These results are consistent with each other within the errors from $t/a = 11$ to $14$,
and it approaches to 180$^\circ$ at $k \rightarrow 0$. 
As shown in Fig.~\ref{fig:NOmegaPhaseShift}~(Right), the scattering length $a_0 \equiv - \lim_{k\rightarrow 0}
\tan\delta_0(k)/k$ becomes positive.
These results mean the formation of a quasi-bound state of $N\Omega$ in the $^5$S$_2$ channel.

The effective range expansion up to the NLO
is given by
  $k\cot\delta_0 = - \frac{1}{a_0} + \frac{1}{2} r_\mathrm{eff}k^2$,
where $r_\mathrm{eff}$ is the effective range.
These parameters are found to be
$a_0 = 5.30(0.44)(^{+0.16}_{-0.01})$~fm and
$r_\mathrm{eff} = 1.26(0.01)(^{+0.02}_{-0.01})$~fm,
where the central values and the statistical errors are obtained at $t/a = 12$,
and the systematic errors in the second parentheses
are estimated from the central values at $t/a = 11$, 13 and 14.
The smallness of $r_\mathrm{eff}/a_0$ suggests
this dibaryon system is close to the unitary limit.

The binding energy $B$ and the root mean square distance $\sqrt{\langle r^2 \rangle}$
are found to be
  $B = 1.54(0.30)(^{+0.04}_{-0.10})$~MeV
  and $\sqrt{\langle r^2\rangle} = 3.77(0.31)(^{+0.11}_{-0.01})$~fm.
The size of the $N\Omega$($^5$S$_2$) bound state
is comparable to its scattering length, which implies 
this system is a loosely bound state
like deuteron from the experiments
and $\Omega\Omega$($^1$S$_0$) from the lattice QCD calculation~\cite{Gongyo:2017fjb}.

Finally, we consider the $p\Omega^{-}$($^5$S$_2$) system with the Coulomb interaction.
By using $V_\mathrm{fit}(r) - \alpha/r$ with $\alpha \equiv e^2/(4\pi) = 1/137.036$, we obtain
$B_{p\Omega^{-}} = 2.46(0.34)(^{+0.04}_{-0.11})$~MeV
and $\sqrt{\langle r^2\rangle}_{p\Omega^{-}} = 3.24(0.19)(^{+0.06}_{-0.00})$~fm.
Due to the additional attractive interaction,
$p\Omega^{-}$($^5$S$_2$) dibaryon becomes slightly deeper and compact bound system than $n\Omega^{-}$($^5$S$_2$) dibaryon.

\section{Summary}

In this paper, we have discussed the two-baryon systems from lattice QCD
by using the direct method and the HAL QCD method.
The contamination of the scattering states can cause serious uncertainties 
in the eigenenergies from the simple plateau fitting in the direct method.
Therefore, the variational method is mandatory to extract reliable eigenenergies.
On the other hand, the HAL QCD method is free from such a problem,
and systematic uncertainty in the derivative expansion has been shown to be under control.

By using the HAL QCD method, we have studied
the $N\Omega$($^5$S$_2$) system with almost physical quark masses.
We have found a strong attractive potential in all distances,
which produces a dibaryon state.
This state can be searched by the two-particle correlation
at the heavy ion collisions~\cite{Morita:2016auo}.
Recently, $N\Omega$ correlation is reported by the STAR Collaboration at RHIC~\cite{STAR:2018uho},
and updated theoretical analyses by using the HAL QCD potentials
near the physical point will be reported elsewhere~\cite{Morita:2018}.

\section*{Acknowledgements}

The lattice QCD calculations have been performed on Blue Gene/Q at KEK 
(Nos. 12/13-19, 13/14-22, 14/15-21, 15/16-12),
HA-PACS at University of Tsukuba (Nos. 13a-23, 14a-20)
and K computer at AICS (hp120281, hp130023, hp140209, hp150085, hp150223,
hp150262, hp160093, hp160211, hp170230),
HOKUSAI FX100 computer at RIKEN, Wako (G15023, G16030, G17002).
This research was supported by MEXT as ``Priority Issue on Post-K computer''
(Elucidation of the Fundamental Laws and Evolution of the Universe)
and JICFuS.


\begin{thebibliography}{99}

\bibitem{Jaffe:1976yi} 
  R.~L.~Jaffe,
  Phys.\ Rev.\ Lett.\  {\bf 38}, 195 (1977)
  Erratum: [Phys.\ Rev.\ Lett.\  {\bf 38}, 617 (1977)].
  10.1103/PhysRevLett.38.195

\bibitem{Inoue:2011ai} 
  T.~Inoue {\it et al.} [HAL QCD Collaboration],
  Nucl.\ Phys.\ A {\bf 881}, 28 (2012)
  [arXiv:1112.5926 [hep-lat]].

\bibitem{Sasaki:2018mzh}
  K.~Sasaki {\it et al.} [HAL QCD Collaboration],
  EPJ Web Conf.\  {\bf 175}, 05010 (2018).


\bibitem{Francis:2018qch} 
  A.~Francis, J.~R.~Green, P.~M.~Junnarkar, C.~Miao, T.~D.~Rae and H.~Wittig,
  arXiv:1805.03966 [hep-lat].

\bibitem{Goldman:1987ma} 
  J.~T.~Goldman, K.~Maltman, G.~J.~Stephenson, Jr., K.~E.~Schmidt and F.~Wang,
  Phys.\ Rev.\ Lett.\  {\bf 59}, 627 (1987).

\bibitem{Oka:1988yq} 
  M.~Oka,
  Phys.\ Rev.\ D {\bf 38}, 298 (1988).

\bibitem{Li:1999bc} 
  Q.~B.~Li and P.~N.~Shen,
  Eur.\ Phys.\ J.\ A {\bf 8}, 417 (2000)
  10.1007/s10050-000-5080-y
  [nucl-th/9910060].

\bibitem{Pang:2003ty} 
  H.~r.~Pang, J.~l.~Ping, F.~Wang, J.~T.~Goldman and E.~g.~Zhao,
  Phys.\ Rev.\ C {\bf 69}, 065207 (2004)
  [nucl-th/0306043].

\bibitem{Zhu:2015sna} 
  X.~Zhu, H.~Huang, J.~Ping and F.~Wang,
  Phys.\ Rev.\ C {\bf 92}, no. 3, 035210 (2015)
  [arXiv:1507.05851 [hep-ph]].

\bibitem{Huang:2015yza} 
  H.~Huang, J.~Ping and F.~Wang,
  Phys.\ Rev.\ C {\bf 92}, 065202 (2015)
  [arXiv:1507.07124 [hep-ph]].

\bibitem{Sekihara:2018tsb} 
  T.~Sekihara, Y.~Kamiya and T.~Hyodo,
  Phys.\ Rev.\ C {\bf 98}, 015205 (2018)
  [arXiv:1805.04024 [hep-ph]].

\bibitem{Etminan:2014tya} 
  F.~Etminan {\it et al.} [HAL QCD Collaboration],
  Nucl.\ Phys.\ A {\bf 928}, 89 (2014)
  [arXiv:1403.7284 [hep-lat]].


\bibitem{Davoudi:2017ddj} 
  Z.~Davoudi,
  EPJ Web Conf.\  {\bf 175}, 01022 (2018)
  [arXiv:1711.02020 [hep-lat]],
  and the references therein.

\bibitem{Ishii:2006ec}
  N.~Ishii, S.~Aoki and T.~Hatsuda,
  Phys.\ Rev.\ Lett.\  {\bf 99} (2007) 022001 [nucl-th/0611096].


\bibitem{Aoki:2012tk}
  S.~Aoki {\it et al.} [HAL QCD Collaboration],
  PTEP {\bf 2012} (2012) 01A105 [arXiv:1206.5088 [hep-lat]].



\bibitem{HALQCD:2012aa}
  N.~Ishii {\it et al.} [HAL QCD Collaboration],
  Phys.\ Lett.\ B {\bf 712} (2012) 437  [arXiv:1203.3642 [hep-lat]].


\bibitem{Iritani:2016jie}
  T.~Iritani {\it et al.},
  JHEP {\bf 1610} (2016) 101
  [arXiv:1607.06371 [hep-lat]].

\bibitem{Iritani:2018zbt} 
  T.~Iritani {\it et al.} [HAL QCD Collaboration],
  arXiv:1805.02365 [hep-lat].

\bibitem{Iritani:2018sra} 
  T.~Iritani {\it et al.},
  arXiv:1810.03416 [hep-lat].

\bibitem{Iritani:2017rlk} 
  T.~Iritani {\it et al.},
  Phys.\ Rev.\ D {\bf 96}, no. 3, 034521 (2017)
  [arXiv:1703.07210 [hep-lat]].


\bibitem{Aoki:2017byw} 
  S.~Aoki, T.~Doi and T.~Iritani,
  EPJ Web Conf.\  {\bf 175}, 05006 (2018)
  [arXiv:1707.08800 [hep-lat]].

\bibitem{Luscher:1991} 
  M.~L\"uscher, Nucl. Phys. B {\bf 354}, 531 (1991).

\bibitem{Ishikawa:2015rho} 
  K.-I.~Ishikawa {\it et al.} [PACS Collaboration],
  PoS LATTICE {\bf 2015}, 075 (2016)
  [arXiv:1511.09222 [hep-lat]].

\bibitem{Gongyo:2017fjb} 
  S.~Gongyo {\it et al.},
  Phys.\ Rev.\ Lett.\  {\bf 120}, no. 21, 212001 (2018)
  [arXiv:1709.00654 [hep-lat]].

\bibitem{Morita:2016auo} 
  K.~Morita, A.~Ohnishi, F.~Etminan and T.~Hatsuda,
  Phys.\ Rev.\ C {\bf 94}, no. 3, 031901 (2016)
  [arXiv:1605.06765 [hep-ph]].

\bibitem{STAR:2018uho} 
  J.~Adam {\it et al.} [STAR Collaboration],
  arXiv:1808.02511 [hep-ex].

\bibitem{Morita:2018}
  K.~Morita, S.~Gongyo, T.~Hatsuda, T.~Iritani, A.~Ohnishi, and K.~Sasaki,
  in preparation.


\end{thebibliography}
\end{document}